\documentclass[11pt]{article}

\usepackage[margin=1in]{geometry}
\usepackage{amsmath,amsfonts,amssymb,amsthm,mathtools,nicefrac,mymath}
\newtheorem{definition}{Definition}
\usepackage{nicematrix}
\NiceMatrixOptions{
  code-for-first-row = \scriptstyle,
  code-for-first-col = \scriptstyle,
}
\usepackage[algo2e,ruled]{algorithm2e}
\usepackage{listings}
\usepackage{caption}
\usepackage{subcaption}
\usepackage{nicefrac}
\usepackage{tabularx}

\usepackage[braket]{qcircuit}
\usepackage{environ}
\newcommand{\myqctmp}[2][0.25]{\Qcircuit @C=#2em @R=#1em @!R}
\NewEnviron{myqcircuit}[1][0.25]{\vcenter{\myqctmp[#1]{0.5} {\BODY}}}
\NewEnviron{myqcircuit*}[2]{\vcenter{\myqctmp[#1]{#2} {\BODY}}}

\definecolor{myblue}{rgb}{0,0.4470,0.7410}
\definecolor{myred}{rgb}{0.8500,0.3250,0.0980}
\definecolor{myorange}{rgb}{0.9290,0.6940,0.1250}
\definecolor{mypurple}{rgb}{0.4940,0.1840,0.5560}
\definecolor{mygreen}{rgb}{0.4660,0.6740,0.1880}
\definecolor{mylightblue}{rgb}{0.3010,0.7450,0.9330}
\definecolor{mydarkred}{rgb}{0.6350,0.0780,0.1840}

\usepackage{tikz}
\usepackage{pgfplots}
\pgfplotsset{
  compat=newest,
  table/header=false,
  tick label style={font=\scriptsize},
  label style={font=\scriptsize},
  legend style={font=\scriptsize},
  legend cell align=left,
  colormap={parula}{
    rgb255=(53,42,135)
    rgb255=(15,92,221)
    rgb255=(18,125,216)
    rgb255=(7,156,207)
    rgb255=(21,177,180)
    rgb255=(89,189,140)
    rgb255=(165,190,107)
    rgb255=(225,185,82)
    rgb255=(252,206,46)
    rgb255=(249,251,14)
  }
}
\pgfplotsset{
  myColOne/.style={myblue},
  myColTwo/.style={myred},
  myColThr/.style={myorange},
  myColFou/.style={mypurple},
  myColFiv/.style={mygreen},
  myColSix/.style={mylightblue},
  myColSev/.style={mydarkred}
}
\pgfplotsset{
  myStyOne/.style={myColOne,thick,mark=+},
  myStyTwo/.style={myColTwo,thick,mark=+},
  myStyThr/.style={myColThr,thick,mark=+},
  myStyFou/.style={myColFou,thick,mark=+},
  myStyFiv/.style={myColFiv,thick,mark=+},
  myStySix/.style={myColSix,thick,mark=+},
  myStySev/.style={myColSev,thick,mark=+},
  myStyRes/.style={black,densely dotted},
}
\pgfplotsset{
  myStyOn/.style={myblue,     thick,mark=asterisk},
  myStyTw/.style={myred,      thick,mark=asterisk},
  myStyTh/.style={myorange,   thick,mark=asterisk},
  myStyFo/.style={mypurple,   thick,mark=asterisk},
  myStyFi/.style={mygreen,    thick,mark=asterisk},
  myStySi/.style={mylightblue,thick,mark=asterisk},
  myStySe/.style={mydarkred,  thick,mark=asterisk},
}
\pgfplotsset{
  every axis/.append style={
    label style={font=\footnotesize},
  }
}
\newcommand{\datfile}[1]{dat/#1.dat}

\newcommand{\starA}{\textcolor{black}{\star}}
\newcommand{\starB}{\textcolor{blue}{\star}}
\newcommand{\starC}{\textcolor{red}{\star}}
\newcommand{\starD}{\textcolor{cyan}{\star}}

\newcommand{\qclabpp}{\texttt{qclab++}}
\newcommand{\cpp}{\texttt{C++}}
\newcommand{\qclab}{\texttt{qclab}}
\newcommand{\matlab}{\textsc{Matlab}}
\newcommand{\openmp}{\texttt{OpenMP}}

\usepackage[colorlinks,urlcolor=blue,linkcolor=blue,citecolor=blue]{hyperref}
\usepackage[capitalize,nameinlink]{cleveref}

\title{\bf QCLAB++: Simulating Quantum Circuits on GPUs}
\author{%
Roel Van Beeumen\textsuperscript{1},
Daan Camps\textsuperscript{2},
Neil Mehta\textsuperscript{2}
}
\date{\scriptsize%
\textsuperscript{1}Applied Mathematics and Computational Research Division, Lawrence Berkeley National Laboratory, Berkeley, CA, USA\\
\textsuperscript{2}National Energy Research Scientific Computing Center, Lawrence Berkeley National Laboratory, Berkeley, CA, USA\\
}

\begin{document}

\maketitle

\begin{abstract}
We introduce \qclabpp, a light-weight, fully-templated \cpp\ package for GPU-accelerated quantum circuit simulations.
The code offers a high degree of portability as it has no external dependencies and the GPU kernels are generated through \openmp\ offloading.
\qclabpp\ is designed for performance and numerical stability through highly optimized gate simulation algorithms for 1-qubit, controlled 1-qubit, and 2-qubit gates.
Furthermore, we also introduce \qclab\, a quantum circuit toolbox for \matlab\ with a syntax that mimics \qclabpp.
This provides users the flexibility and ease of use of a scripting language like \matlab\ for studying their quantum algorithms, while offering high-performance
GPU acceleration when required.
As such, the \qclabpp\ library offers a unique combination of features.
We compare the CPU simulator in \qclabpp\ with the GPU kernels generated by \openmp\ and observe a speedup of over $40\times$.
Furthermore, we also compare \qclabpp\ to other circuit simulation packages, such as cirq-qsim and qibo, in a series of benchmarks conducted on NERSC's Perlmutter system and illustrate its competitiveness.
\end{abstract}

\section{Introduction}
\label{sec:intro}

The field of quantum computing is quickly progressing as is illustrated by recent developments in quantum hardware.
Recent initial steps towards experimental realization of error correcting schemes in quantum processors based on
superconducting~\cite{GoogleErrorCorrection} and ion traps~\cite{QuantinuumErrorSuppression} are encouraging milestones for the field.
Other qubit modalities, such as neutral atoms~\cite{Williams2022}, have also shown significant promise.

Most quantum hardware platforms that are currently being developed support a circuit model of quantum computation.
It is thus of interest to the community of quantum algorithm researchers to have access to advanced computational tools that can help them in their research
while quantum hardware is being developed.
In addition to access to QPUs, quantum circuit simulators are of interest for a variety of reasons. First, it allows the researcher to
prototype and investigate novel algorithms without using up valuable quantum resources for debugging and testing their ideas.
This can instead all be done using simulations on classical hardware.
Second, as the currently available quantum hardware is still suffering from noise,
it is not possible to study algorithms that require a circuit depth that lies beyond the coherence limits of the device.
Many algorithms of interest, for example in molecular sciences~\cite{Liu2022}, require deep circuits.

This has sparked the development of multiple quantum simulators each with their own advantages and strengths.
The \textit{Cirq} toolbox~\cite{cirq} with the \textit{qsim} simulator~\cite{qsim} is developed by Google Quantum AI,
IBM has \textit{Qiskit}~\cite{Qiskit} with the \textit{Aer} simulator backend, and \textit{PennyLane}~\cite{pennylane} is
a differentiable quantum programming framework from Xanadu. Similar efforts are pursued in \textit{Qibo}~\cite{qibo_paper,qibojit_paper},
which includes a JIT compiler for quantum simulation, and \textit{Quest}~\cite{quest}.
All of the aforementioned packages, with the exception of Quest, provide a Python API to generate the quantum program.
Recently, NVIDIA has introduced their \textit{cuQuantum} SDK~\cite{cuquantum} which offers a collection of GPU kernels to perform quantum gate simulations.
Many existing quantum simulation packages, such as Cirq, Qiskit, PennyLane and Qibo, have started to support the cuQuantum SDK to run circuit simulations
on NVIDIA hardware.

In this work, we introduce \emph{two} alternative quantum circuit programming and simulation frameworks, called \qclabpp\footnote{\qclabpp: \url{https://github.com/QuantumComputingLab/qclabpp}}~\cite{qclabpp} and \qclab\footnote{\qclab: \url{https://github.com/QuantumComputingLab/qclab}}~\cite{qclab}.
The former is a light-weight fully-templated \cpp\ library that supports GPU accelerated state vector simulations through \openmp\ offloading,
while the latter is a \matlab\ toolbox that has an accessible interface that makes use of the advantages the \matlab\ IDE offers. 
Both libraries provide a similar programming interface, see \Cref{fig:MatlabvCpp}, which lowers the barrier for converting simulation codes
from one framework to the other and allows for a streamlined workflow where ideas are first prototyped in \matlab\ and later simulated at larger 
scale on the GPU with \qclabpp.
As an example, we illustrate this for the QFT circuit~\cite{qft, qft_camps} in \cref{fig:MatlabvCpp}.
Remark that both implementations are very similar and easy to convert into each other.

\begin{figure}[hbtp]
\centering
\begin{minipage}{0.45\textwidth}
\centering
\begin{lstlisting}%
[language=Matlab,basicstyle=\ttfamily\scriptsize,keywordstyle=\color{blue},commentstyle=\color{green!70!black}]
% number of qubits
n = 5;

% quantum circuit
circ = qclab.QCircuit(n);

% blocks
for i=0:n-1
  % Hadamard
  circ.push_back( H(i) );
  % diagonal blocks
  for j=2:n-i
    ctrl = j+i-1;
    th = -2*pi/2^j;
    circ.push_back( ...
      CP(ctrl, i, 1, th) );
  end
end
  
% swaps
for i=0:floor(n/2)-1
  circ.push_back( ...
    SWAP(i, n-i-1) );
end
\end{lstlisting}%
\end{minipage}%
\hfill%
\begin{minipage}{0.5\textwidth}
\centering
\begin{lstlisting}[language=C++,basicstyle=\ttfamily\scriptsize,keywordstyle=\color{blue},commentstyle=\color{green!70!black}]
// number of qubits
int n = 5 ;

// quantum circuit
qclab::QCircuit<std::complex<double>> circ(n);

// blocks
for (int i=0; i<n; i++) {
  // Hadamard
  circ.push_back( std::make_unique<H>(i) );
  // diagonal blocks
  for (int j=2; j<=n-i; j++) {
    int ctrl = j+i-1;
    double th = -2*pi/(1<<j);
    circ.push_back(
      std::make_unique<CP>(ctrl, i, th) );
  }
}

// swaps
for (int i=0; i<n/2; i++) {
  circ.push_back(
    std::make_unique<SWAP>(i, n-i-1) );
}
\end{lstlisting}
\end{minipage}%
\caption{\matlab\ implementation of the QFT circuit~\cite{qft, qft_camps} with the \qclab\ toolbox (\emph{left}) and equivalent \cpp\ implementation using \qclabpp\ (\emph{right}).\label{fig:MatlabvCpp}}
\end{figure}

Our simulators are \emph{strong} simulators that keep track of all amplitudes in the wavefunction as it is modified by the quantum circuit.
This simulation problem naturally scales exponentially in the number of qubits of the quantum system and linearly in the number of gates in the circuit.
The main advantages of strong simulation is that we have access to every amplitude at each step of the simulation and that all computations are exact up to
numerical precision. The main disadvantage is that the wavefunction scales exponentially and this imposes a limit on the size of system one can simulate classically. This limit is in the range of 40-50 qubits.
Approximate simulation methods using tensor network or sparse representation of the wave function can scale to larger systems but cannot approximate every state
in Hilbert space up to the same accuracy.

The remainder of this paper is organized as follows.
In \cref{sec:eff}, we describe the optimized state vector simulator algorithms for 1-qubit, controlled 1-qubit, and 2-qubit gates.
In \cref{sec:impl}, we discuss the implementation details and the \openmp\ offloading used to accelerate \qclabpp.
In \cref{sec:exp}, we compare the CPU and GPU implementations of \qclabpp\ and benchmark it with other GPU-accelerated state vector simulation packages.

\section{Efficient quantum gate simulations}
\label{sec:eff}

\begin{figure}[hbtp]
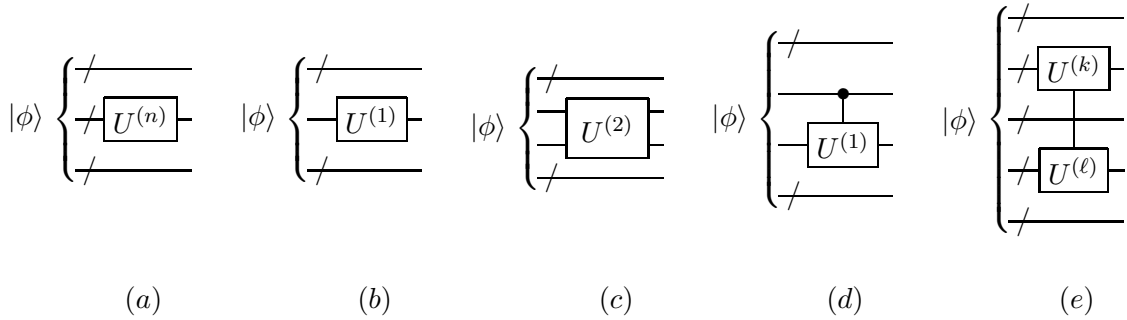

\centering
\begin{align*}
\quad{\begin{myqcircuit}
 & {/} \qw & \qw \ & \qw \\
 & {/} \qw & \gate{U\p{n}} & \qw \\
 & {/} \qw & \qw  & \qw \\
{\inputgroupv{1}{3}{0.7em}{1.7em}{\ket{\phi}}} \\
\end{myqcircuit}}
&&
\quad{\begin{myqcircuit}
 & {/} \qw & \qw \ & \qw \\
 & \qw & \gate{U\p{1}} & \qw \\
 & {/} \qw & \qw  & \qw \\
{\inputgroupv{1}{3}{0.7em}{1.7em}{\ket{\phi}}} \\
\end{myqcircuit}}
&&
\quad{\begin{myqcircuit}
 & {/} \qw & \qw \ & \qw \\
 & \qw & \multigate{1}{U\p{2}} & \qw \\
 & \qw & \ghost{U\p{2}} & \qw \\
 & {/} \qw & \qw  & \qw \\
{\inputgroupv{1}{4}{0.7em}{1.7em}{\ket{\phi}}} \\
\end{myqcircuit}}
&&
\quad{\begin{myqcircuit}
 & {/} \qw & \qw \ & \qw \\
 & \qw & \ctrl{1} & \qw \\
 & \qw & \gate{U\p{1}} & \qw \\
 & {/} \qw & \qw  & \qw \\
{\inputgroupv{1}{4}{0.7em}{2.6em}{\ket{\phi}}} \\
\end{myqcircuit}}
&&
\quad{\begin{myqcircuit}
 & {/} \qw & \qw \ & \qw \\
 & {/} \qw & \gate{U\p{k}}\qwx[2] & \qw \\
 & {/} \qw & \qw & \qw \\
 & {/} \qw & \gate{U\p{\ell}} & \qw \\
 & {/} \qw & \qw  & \qw \\
{\inputgroupv{1}{5}{0.7em}{3.6em}{\ket{\phi}}} \\
\end{myqcircuit}}\\
(a)\quad &&
(b)\quad &&
(c)\quad &&
(d)\quad &&
(e)\quad
\end{align*}
\caption{Application of (a) a dense $n$-qubit quantum gate, (b) a $1$-qubit quantum gate, (c) a $2$-qubit quantum gate, (d) a controlled $1$-qubit quantum gate, and (e) a $k+\ell$ quantum gate acting on noncontiguous qubits to a quantum state $\ket\phi$.}
\label{fig:qgates}
\end{figure}

The application of a unitary quantum gate $U$ to a given state $\ket\phi \inC[2^n]$  (\cref{fig:qgates}) corresponds mathematically to the matrix-vector multiplication (\emph{matvec})
\begin{equation}
\ket\psi = (\eye_l \otimes U \otimes \eye_r) \ket\phi,
\label{eq:kron}
\end{equation}
where $\otimes$ denotes the Kronecker product and $\eye_l,\eye_r$ are identity matrices of appropriate dimensions.
A naive implementation of quantum circuit simulator would carry out the Kronecker products of \eqref{eq:kron} explicitly, and require 2 nested for loops.

In this section, we introduce more efficient algorithms for quantum gate simulation that only consist of 1 simple for loop combined with bit operations for index calculations.
We therefore will use the \emph{big-endian} binary convention as given by the following definition.

\begin{definition}[Binary representation]
\label{def:bin}
We define the binary representation of $j \inN: 0 \leq j \leq 2^n-1$ as follows
\[
j = [\ttj_0 \ttj_1 \cdots \ttj_{n-1}]
  = \ttj_0 \cdot 2^{n-1} + \ttj_1 \cdot 2^{n-2} + \cdots +
    \ttj_{n-1} \cdot 2^0,
\]
where $\ttj_i \in \lbrace 0,1 \rbrace$ for $i = 0,\ldots,n-1$.
\end{definition}

\subsection{1-qubit gates}
\label{sec:1q}

\begin{figure}[tbp]
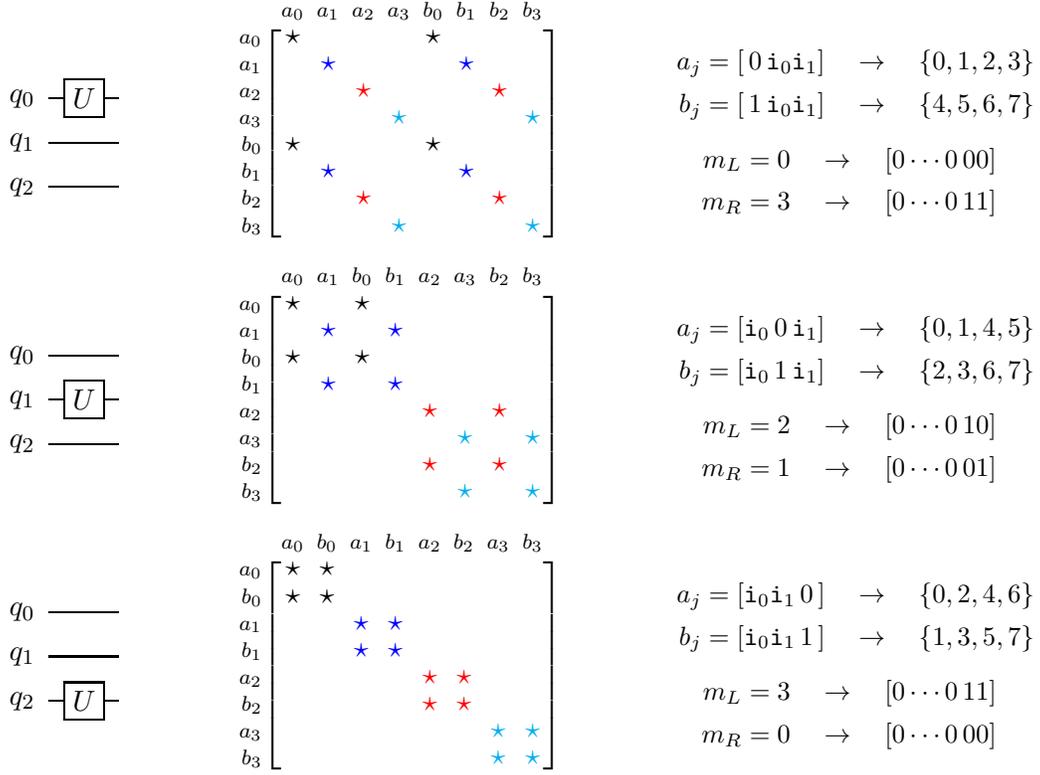

\centering
\begin{align*}
\quad{\begin{myqcircuit}
\lstick{q_0} & \gate{U} & \qw \\
\lstick{q_1} & \qw      & \qw \\
\lstick{q_2} & \qw      & \qw
\end{myqcircuit}} &&
\setlength{\arraycolsep}{0.5ex}\renewcommand{\arraystretch}{0.75}
\begin{bNiceMatrix}[first-row,first-col]
      & a_0 & a_1 & a_2 & a_3 & b_0 & b_1 & b_2 & b_3 \\
  a_0 & \starA & & & & \starA \\
  a_1 & & \starB & & & & \starB \\
  a_2 & & & \starC & & & & \starC \\
  a_3 & & & & \starD & & & & \starD \\
  b_0 & \starA & & & & \starA \\
  b_1 & & \starB & & & & \starB \\
  b_2 & & & \starC & & & & \starC \\
  b_3 & & & & \starD & & & & \starD \\
\end{bNiceMatrix} &&
{\small\begin{aligned}
a_j &= [\, 0 \, \tti_0 \tti_1] \quad \to \quad \{ 0,1,2,3 \} \\
b_j &= [\, 1 \, \tti_0 \tti_1] \quad \to \quad \{ 4,5,6,7 \} \\[5pt]
    & \begin{aligned} m_L &= 0 \quad \to \quad [0 \cdots 0 \, 0 0] \\
                      m_R &= 3 \quad \to \quad [0 \cdots 0 \, 1 1] \\
      \end{aligned} \\
\end{aligned}} \\[5pt]
{\begin{myqcircuit}
\lstick{q_0} & \qw      & \qw \\
\lstick{q_1} & \gate{U} & \qw \\
\lstick{q_2} & \qw      & \qw
\end{myqcircuit}} &&
\setlength{\arraycolsep}{0.5ex}\renewcommand{\arraystretch}{0.75}
\begin{bNiceMatrix}[first-row,first-col]
      & a_0 & a_1 & b_0 & b_1 & a_2 & a_3 & b_2 & b_3 \\
  a_0 & \starA & & \starA \\
  a_1 & & \starB & & \starB \\
  b_0 & \starA & & \starA \\
  b_1 & & \starB & & \starB \\
  a_2 & & & & & \starC & & \starC \\
  a_3 & & & & & & \starD & & \starD \\
  b_2 & & & & & \starC & & \starC \\
  b_3 & & & & & & \starD & & \starD \\
\end{bNiceMatrix} &&
{\small\begin{aligned}
a_j &= [\tti_0 \, 0 \, \tti_1] \quad \to \quad \{ 0,1,4,5 \} \\
b_j &= [\tti_0 \, 1 \, \tti_1] \quad \to \quad \{ 2,3,6,7 \} \\[5pt]
    & \begin{aligned} m_L &= 2 \quad \to \quad [0 \cdots 0 \, 1 0] \\
                      m_R &= 1 \quad \to \quad [0 \cdots 0 \, 0 1] \\
      \end{aligned} \\
\end{aligned}} \\[5pt]
{\begin{myqcircuit}
\lstick{q_0} & \qw      & \qw \\
\lstick{q_1} & \qw      & \qw \\
\lstick{q_2} & \gate{U} & \qw
\end{myqcircuit}} &&
\setlength{\arraycolsep}{0.5ex}\renewcommand{\arraystretch}{0.75}
\begin{bNiceMatrix}[first-row,first-col]
      & a_0 & b_0 & a_1 & b_1 & a_2 & b_2 & a_3 & b_3 \\
  a_0 & \starA & \starA \\
  b_0 & \starA & \starA \\
  a_1 & & & \starB & \starB \\
  b_1 & & & \starB & \starB \\
  a_2 & & & & & \starC & \starC \\
  b_2 & & & & & \starC & \starC \\
  a_3 & & & & & & & \starD & \starD \\
  b_3 & & & & & & & \starD & \starD \\
\end{bNiceMatrix} &&
{\small\begin{aligned}
a_j &= [\tti_0 \tti_1 \, 0 \,] \quad \to \quad \{ 0,2,4,6 \} \\
b_j &= [\tti_0 \tti_1 \, 1 \,] \quad \to \quad \{ 1,3,5,7 \} \\[5pt]
    & \begin{aligned} m_L &= 3 \quad \to \quad [0 \cdots 0 \, 1 1] \\
                      m_R &= 0 \quad \to \quad [0 \cdots 0 \, 0 0] \\
      \end{aligned} \\
\end{aligned}}
\end{align*}
\caption{Graphical illustration of applying a 1-qubit gate $U$.}
\label{fig:1q}
\end{figure}

We start with the simplest class of quantum gates that only act on a single qubit
\begin{equation}
U := \begin{bmatrix} u_{0,0} & u_{0,1} \\ u_{1,0} & u_{1,1} \end{bmatrix}.
\label{eq:U1}
\end{equation}
As illustrated by the different colors in \cref{fig:1q}, the state vector simulation of a 1-qubit gate breaks down in a series of $2 \times 2$ matvec operations
\begin{equation}
\begin{bmatrix}
\psi_{\tti_0 \tti_1 \cdots \tti_{q-1} \, 0 \, \tti_{q+1} \cdots \tti_{n-1}} \\
\psi_{\tti_0 \tti_1 \cdots \tti_{q-1} \, 1 \, \tti_{q+1} \cdots \tti_{n-1}} \\
\end{bmatrix}
=
\begin{bmatrix} u_{0,0} & u_{0,1} \\ u_{1,0} & u_{1,1} \end{bmatrix}
\begin{bmatrix}
\phi_{\tti_0 \tti_1 \cdots \tti_{q-1} \, 0 \, \tti_{q+1} \cdots \tti_{n-1}} \\
\phi_{\tti_0 \tti_1 \cdots \tti_{q-1} \, 1 \, \tti_{q+1} \cdots \tti_{n-1}} \\
\end{bmatrix},
\label{eq:1q-matvec}
\end{equation}
where $q$ is the qubit the gate $U$ is acting on.
Depending on $q$, the indices in \eqref{eq:1q-matvec}
\begin{align}
\begin{aligned}
[\tti_0 \tti_1 \cdots \tti_{q-1} \, 0 \, \tti_{q+1} \cdots \tti_{n-1}] &=: a_j,\\
[\tti_0 \tti_1 \cdots \tti_{q-1} \, 1 \, \tti_{q+1} \cdots \tti_{n-1}] &=: b_j,
\end{aligned} && j = 0,1,\ldots,2^{n-1}-1,
\label{eq:1q-aj-bj}
\end{align}
can be next to each other or separated up to $2^{n-1}$.
We will now describe an efficient way to calculate \eqref{eq:1q-aj-bj} solely based on bit operations.

First, we observe that the binary representations of $a_j$ and $b_j$ can be split into 3 parts, i.e., the bits left of 0 (and 1), the 0 (and 1) bit, and the bits right of 0 (and 1), respectively.
Without loss of generality, we can select the bits $\tti_\ell$ for $0 \leq \ell \leq n-1$ and $\ell \neq q$, so that
\begin{equation}
j = [\tti_0 \tti_1 \cdots \tti_{q-1} \,|\, \tti_{q+1} \cdots \tti_{n-1}].
\end{equation}
Next, using the following left and right bit masks
\begin{align}
m_L &:= [ 0 \cdots 0 \,| \overbrace{1 \cdots 1}^q |\, 0 \cdots 0 ]
      = \left( 2^{n-1} - 1 \right) - \left( 2^{n-q-1} - 1 \right), \\
m_R &:= [ 0 \cdots 0 \,|\, 0 \cdots 0 \,| \underbrace{1 \cdots 1}_{n-q-1} ]
      = 2^{n-q-1} - 1,
\end{align}
we can compute each $a_j$ as the sum of $j$ masked with $m_R$ and a left shift of $j$ masked with $m_L$,
\begin{equation}
a_j = (j \,\&\, m_R) + [(j \,\&\, m_L) \ll 1],
\end{equation}
and the corresponding $b_j$ index results from the following sum
\begin{equation}
b_j = a_j + 2^{n-q-1}.
\end{equation}

The algorithm for the vector simulation of a general 1-qubit gate is given in \cref{alg:1q}.
Note that for 1-qubit gates with some of the elements in \eqref{eq:U1} equal to 0, \cref{alg:1q-matvec1,alg:1q-matvec2} can be simplified.
For example, the Pauli-$X$ gate is just a swap operation
\begin{align*}
\psi[a_j] &= \phi[b_j], \\
\psi[b_j] &= \phi[a_j],
\end{align*}
and the Pauli-$Y$ gate
\begin{align*}
\psi[a_j] &= -i \phi[b_j], \\
\psi[b_j] &= \phantom{-}i \phi[a_j].
\end{align*}
Remark that for the Pauli-$Z$ gate only half of the vector elements ($b_j$) need to be updated
\begin{align*}
\psi[a_j] &= \phantom{-}\phi[a_j], \\
\psi[b_j] &= -\phi[b_j].
\end{align*}

\begin{algorithm2e}[hbtp]
\caption{Apply 1-qubit gate}
\label{alg:1q}
\vspace{5pt}
\KwIn{Input state $\ket\phi$.}
\KwOut{Output state $\ket\psi = U \ket\phi$.}
\vspace{5pt}
\nl Right mask: $m_R = 2^{n-q-1} - 1$ \\
\nl Left mask:  $m_L = 2^{n-1} - 1 - m_R$ \\
\For{$j=0,1,\ldots,2^{n-1}-1$}{
\vspace{1pt}
\nl $a_j = (j \,\&\, m_R) + [(j \,\&\, m_L) \ll 1]$ \\
\nl $b_j = a_j + 2^{n-q-1}$ \\
\stepcounter{AlgoLine}\lnlset{alg:1q-matvec1}{\theAlgoLine}%
    $\psi[a_j] = u_{0,0} \phi[a_j] + u_{0,1} \phi[b_j]$ \\
\stepcounter{AlgoLine}\lnlset{alg:1q-matvec2}{\theAlgoLine}%
    $\psi[b_j] = u_{1,0} \phi[a_j] + u_{1,1} \phi[b_j]$
}
\end{algorithm2e}

\subsection{Controlled 1-qubit gates}
\label{sec:ctrl-1q}

We now consider the controlled 1-qubit gates and distinguish the following 4 different cases
\begin{align}
{\begin{myqcircuit}
\lstick{q_c} & \ctrl1   & \qw \\
\lstick{q_t} & \gate{U} & \qw
\end{myqcircuit}}
&&
{\begin{myqcircuit}
\lstick{q_c} & \ctrlo1  & \qw \\
\lstick{q_t} & \gate{U} & \qw
\end{myqcircuit}}
&&
{\begin{myqcircuit}
\lstick{q_t} & \gate{U}  & \qw \\
\lstick{q_c} & \ctrl{-1} & \qw
\end{myqcircuit}}
&&
{\begin{myqcircuit}
\lstick{q_t} & \gate{U}   & \qw \\
\lstick{q_c} & \ctrlo{-1} & \qw
\end{myqcircuit}}
\label{eq:ctrl-1q-circuits}
\end{align}
where $q_c$ is the control qubit and $q_t$ the target qubit.
As illustrated in \cref{fig:ctrl-1q}, these controlled gates only update half of the elements.
Which elements are updated depend on the actual values of the $q_c$ and $q_t$.

\begin{figure}[t]
\centering
\begin{align*}
\quad{\begin{myqcircuit}
\lstick{q_0} & \ctrl1   & \qw \\
\lstick{q_1} & \gate{U} & \qw \\
\lstick{q_2} & \qw      & \qw
\end{myqcircuit}} &&
\setlength{\arraycolsep}{0.5ex}\renewcommand{\arraystretch}{0.75}
\begin{bNiceMatrix}[first-row,first-col]
      & & & & & a_0 & a_1 & b_0 & b_1 \\
      & 1 & &  \\
      & & 1 & &  \\
      & & & 1 \\
      & & & & 1 \\
  a_0 & & & & & \starC & & \starC \\
  a_1 & & & & & & \starD & & \starD \\
  b_0 & & & & & \starC & & \starC \\
  b_1 & & & & & & \starD & & \starD \\
\end{bNiceMatrix} &&
{\small\begin{aligned}
a_j &= [\, \underline{\textbf{1}} \, 0 \, \tti_0] \quad \to \quad \{ 4,5 \} \\
b_j &= [\, \underline{\textbf{1}} \, 1 \, \tti_0] \quad \to \quad \{ 6,7 \} \\
\end{aligned}} \\[5pt]
{\begin{myqcircuit}
\lstick{q_0} & \ctrlo1  & \qw \\
\lstick{q_1} & \gate{U} & \qw \\
\lstick{q_2} & \qw      & \qw
\end{myqcircuit}} &&
\setlength{\arraycolsep}{0.5ex}\renewcommand{\arraystretch}{0.75}
\begin{bNiceMatrix}[first-row,first-col]
      & a_0 & a_1 & b_0 & b_1 \\
  a_0 & \starA & & \starA \\
  a_1 & & \starB & & \starB \\
  b_0 & \starA & & \starA \\
  b_1 & & \starB & & \starB \\
      & & & & & 1 \\
      & & & & & & 1 \\
      & & & & & & & 1 \\
      & & & & & & & & 1 \\
\end{bNiceMatrix} &&
{\small\begin{aligned}
a_j &= [\, \underline{\textbf{0}} \, 0 \, \tti_0] \quad \to \quad \{ 0,1 \} \\
b_j &= [\, \underline{\textbf{0}} \, 1 \, \tti_0] \quad \to \quad \{ 2,3 \} \\
\end{aligned}} \\[5pt]
{\begin{myqcircuit}
\lstick{q_0} & \qw       & \qw \\
\lstick{q_1} & \gate{U}  & \qw \\
\lstick{q_2} & \ctrl{-1} & \qw
\end{myqcircuit}} &&
\setlength{\arraycolsep}{0.5ex}\renewcommand{\arraystretch}{0.75}
\begin{bNiceMatrix}[first-row,first-col]
      & & a_0 & & b_0 & & a_1 & & b_1 \\
      & 1 \\
  a_0 & & \starB & & \starB \\
      & & & 1 \\
  b_0 & & \starB & & \starB \\
      & & & & & 1 \\
  a_1 & & & & & & \starD & & \starD \\
      & & & & & & & 1 \\
  b_2 & & & & & & \starD & & \starD \\
\end{bNiceMatrix} &&
{\small\begin{aligned}
a_j &= [\tti_0 \, 0 \, \underline{\textbf{1}} \,] \quad \to \quad \{ 1,5 \} \\
b_j &= [\tti_0 \, 1 \, \underline{\textbf{1}} \,] \quad \to \quad \{ 3,7 \} \\
\end{aligned}} \\[5pt]
{\begin{myqcircuit}
\lstick{q_0} & \qw        & \qw \\
\lstick{q_1} & \gate{U}   & \qw \\
\lstick{q_2} & \ctrlo{-1} & \qw
\end{myqcircuit}} &&
\setlength{\arraycolsep}{0.5ex}\renewcommand{\arraystretch}{0.75}
\begin{bNiceMatrix}[first-row,first-col]
      & a_0 & & b_0 & & a_1 & & b_1 \\
  a_0 & \starA & & \starA \\
      & & 1 \\
  b_0 & \starA & & \starA \\
      & & & & 1 \\
  a_1 & & & & & \starC & & \starC \\
      & & & & & & 1 \\
  b_1 & & & & & \starC & & \starC \\
      & & & & & & & & 1 \\
\end{bNiceMatrix} &&
{\small\begin{aligned}
a_j &= [\tti_0 \, 0 \, \underline{\textbf{0}} \,] \quad \to \quad \{ 0,4 \} \\
b_j &= [\tti_0 \, 1 \, \underline{\textbf{0}} \,] \quad \to \quad \{ 2,6 \} \\
\end{aligned}}
\end{align*}
\caption{Graphical illustration of applying a controlled 1-qubit gate $U$.}
\label{fig:ctrl-1q}
\end{figure}

Similar to the standard 1-qubit gate, the state vector simulation of a controlled 1-qubit gate breaks down in a series of $2 \times 2$ matvec operations
\begin{equation}
\begin{bmatrix}
\psi_{a_j} \\
\psi_{b_j} \\
\end{bmatrix}
=
\begin{bmatrix} u_{0,0} & u_{0,1} \\ u_{1,0} & u_{1,1} \end{bmatrix}
\begin{bmatrix}
\phi_{a_j} \\
\phi_{b_j} \\
\end{bmatrix},
\label{eq:ctrl-1q-matvec}
\end{equation}
where we distinguish between the 2 left cases of \eqref{eq:ctrl-1q-circuits}, i.e., $q_c < q_t$,
\begin{align}
\begin{aligned}
a_j &=
[\tti_0 \tti_1 \cdots \tti_{q_c-1} * \tti_{q_c+1} \cdots \tti_{q_t-1} \, 0 \,
                      \tti_{q_t+1} \cdots \tti_{n-1}], \\
b_j &=
[\tti_0 \tti_1 \cdots \tti_{q_c-1} * \tti_{q_c+1} \cdots \tti_{q_t-1} \, 1 \,
                      \tti_{q_t+1} \cdots \tti_{n-1}],
\end{aligned} && j = 0,1,\ldots,2^{n-2}-1,
\label{eq:ctrl-1q-aj-bj}
\end{align}
and the 2 right cases of \eqref{eq:ctrl-1q-circuits}, i.e., $q_t < q_c$,
\begin{align}
\begin{aligned}
a_j &=
[\tti_0 \tti_1 \cdots \tti_{q_t-1} \, 0 \, \tti_{q_t+1} \cdots \tti_{q_c-1} *
                      \tti_{q_c+1} \cdots \tti_{n-1}], \\
b_j &=
[\tti_0 \tti_1 \cdots \tti_{q_t-1} \, 1 \, \tti_{q_t+1} \cdots \tti_{q_c-1} *
                      \tti_{q_c+1} \cdots \tti_{n-1}],
\end{aligned} && j = 0,1,\ldots,2^{n-2}-1,
\label{eq:ctrl-1q-aj-bj-2}
\end{align}
with $* = 0$ for zero-controlled gates and $* = 1$ for one-controlled gates.
Note that we only have $2^{n-2}$ different values for $j$ in contrast to $2^{n-1}$ values for the standard 1-qubit gate in \eqref{eq:1q-aj-bj}.

Let
\begin{align}
q_0 &= \min( q_c , q_t ), \\
q_1 &= \max( q_c , q_t ).
\end{align}
such that $q_0 < q_1$.
Next, we observe that the indices $a_j$ and $b_j$ in \eqref{eq:ctrl-1q-aj-bj} and \eqref{eq:ctrl-1q-aj-bj-2} can be split into 5 parts.
Without loss of generality, we can select the bits $\tti_\ell$ for $0 \leq \ell \leq n-1$ and $\ell \neq \{q_0,q_1\}$, so that
\begin{equation}
j = [\tti_0 \tti_1 \cdots \tti_{q_0-1} \,|\, \tti_{q_0+1}
                   \cdots \tti_{q_1-1} \,|\, \tti_{q_1+1} \cdots \tti_{n-1}].
\end{equation}
Next, using the following bit masks
\begin{align}
m_L &:= [ 0 \cdots 0 \,| \overbrace{1 \cdots 1}^{q_0} | \!\overbrace{0 \cdots 0}^{q_1-q_0-1}\! |\, 0 \cdots 0 ]
      = \left( 2^{n-2} - 1 \right) - \left( 2^{n-q_0-2} - 1 \right),
      \label{eq:ctrl-1q-mL} \\
m_C &:= [ 0 \cdots 0 \,|\, 0 \cdots 0 \,|\, 1 \cdots 1 \,|\, 0 \cdots 0 ]
      = \left( 2^{n-q_0-2} - 1 \right) - \left( 2^{n-q_1-1} - 1 \right),
      \label{eq:ctrl-1q-mC} \\
m_R &:= [ 0 \cdots 0 \,|\, 0 \cdots 0 \,|\, 0 \cdots 0 \,| \underbrace{1 \cdots 1}_{n-q_1-1} ]
      = 2^{n-q_1-1} - 1, \label{eq:ctrl-1q-mR}
\end{align}
we can compute the zero-controlled $a_j$ indices as the sum of the following 3 terms
\begin{equation}
a_j = (j \,\&\, m_R) + [(j \,\&\, m_C) \ll 1] + [(j \,\&\, m_L) \ll 2],
\label{eq:ctrl-1q-aj}
\end{equation}
and the corresponding $b_j$ indices result from the following sum
\begin{equation}
b_j = a_j + 2^{n-q_t-1}.
\label{eq:ctrl-1q-bj}
\end{equation}
In the case of a one-controlled gate, we need to increment both \eqref{eq:ctrl-1q-aj} and \eqref{eq:ctrl-1q-bj} by $2^{n-q_c-1}$.

The algorithm for the vector simulation of a controlled 1-qubit gate is given in \cref{alg:ctrl-1q}.
Similar to standard 1-qubit gates, controlled 1-qubit gates with some zero elements in \eqref{eq:U1}, \cref{alg:ctrl-1q-matvec1,alg:ctrl-1q-matvec2} can be simplified.
For example, the CNOT gate just swaps half of the elements.

\begin{algorithm2e}[hbtp]
\caption{Apply controlled 1-qubit gate}
\label{alg:ctrl-1q}
\vspace{5pt}
\KwIn{Input state $\ket\phi$.}
\KwOut{Output state $\ket\psi = U \ket\phi$.}
\vspace{5pt}
\nl Right mask:  $m_R = 2^{n-q_1-1} - 1$ \\
\nl Center mask: $m_C = 2^{n-q_0-2} - 1 - m_R$ \\
\nl Left mask:   $m_L = 2^{n-2} - 1 - m_R - m_C$ \\
\For{$j=0,1,\ldots,2^{n-2}-1$}{
\vspace{1pt}
\nl $a_j = (j \,\&\, m_R) + [(j \,\&\, m_C) \ll 1]
                          + [(j \,\&\, m_L) \ll 2]$ \\
\nl $b_j = a_j + 2^{n-q_t-1}$ \\
\If{control state = 1}{
\nl $a_j = a_j + 2^{n-q_c-1}$ \\
\nl $b_j = b_j + 2^{n-q_c-1}$
}
\stepcounter{AlgoLine}\lnlset{alg:ctrl-1q-matvec1}{\theAlgoLine}%
    $\psi[a_j] = u_{0,0} \phi[a_j] + u_{0,1} \phi[b_j]$ \\
\stepcounter{AlgoLine}\lnlset{alg:ctrl-1q-matvec2}{\theAlgoLine}%
    $\psi[b_j] = u_{1,0} \phi[a_j] + u_{1,1} \phi[b_j]$
}
\end{algorithm2e}

\subsection{2-qubit gates}
\label{sec:2q}

The state vector simulation of general 2-qubit gates acting on qubits $q_0$ and $q_1$ breaks down in a series of $4 \times 4$ matvec operations.
For the index calculations we can again make use of the bit mask \eqref{eq:ctrl-1q-mL}--\eqref{eq:ctrl-1q-mR} and the corresponding algorithm is given in \cref{alg:2q}.

\begin{algorithm2e}[hbtp]
\caption{Apply 2-qubit gate}
\label{alg:2q}
\vspace{5pt}
\KwIn{Input state $\ket\phi$.}
\KwOut{Output state $\ket\psi = U \ket\phi$.}
\vspace{5pt}
\nl Right mask:  $m_R = 2^{n-q_1-1} - 1$ \\
\nl Center mask: $m_C = 2^{n-q_0-2} - 1 - m_R$ \\
\nl Left mask:   $m_L = 2^{n-2} - 1 - m_R - m_C$ \\
\For{$j=0,1,\ldots,2^{n-2}-1$}{
\vspace{1pt}
\nl $a_j = (j \,\&\, m_R) + [(j \,\&\, m_C) \ll 1]
                          + [(j \,\&\, m_L) \ll 2]$ \\
\nl $b_j = a_j + 2^{n-q_0-1}$ \\
\nl $c_j = a_j + 2^{n-q_1-1}$ \\
\nl $d_j = a_j + 2^{n-q_0-1} + 2^{n-q_1-1}$ \\
\stepcounter{AlgoLine}\lnlset{alg:2q-matvec1}{\theAlgoLine}%
    $\psi[a_j] = u_{0,0} \phi[a_j] + u_{0,1} \phi[b_j] +
                 u_{0,2} \phi[c_j] + u_{0,3} \phi[d_j]$ \\
\stepcounter{AlgoLine}\lnlset{alg:2q-matvec2}{\theAlgoLine}%
    $\psi[b_j] = u_{1,0} \phi[a_j] + u_{1,1} \phi[b_j] +
                 u_{1,2} \phi[c_j] + u_{1,3} \phi[d_j]$ \\
\stepcounter{AlgoLine}\lnlset{alg:2q-matvec3}{\theAlgoLine}%
    $\psi[c_j] = u_{2,0} \phi[a_j] + u_{2,1} \phi[b_j] +
                 u_{2,2} \phi[c_j] + u_{2,3} \phi[d_j]$ \\
\stepcounter{AlgoLine}\lnlset{alg:2q-matvec4}{\theAlgoLine}%
    $\psi[d_j] = u_{3,0} \phi[a_j] + u_{3,1} \phi[b_j] +
                 u_{3,2} \phi[c_j] + u_{3,3} \phi[d_j]$
}
\end{algorithm2e}

A particular and commonly used 2-qubit gate is the SWAP gate
\begin{equation}
U =
\begin{bmatrix}
1 & 0 & 0 & 0 \\
0 & 0 & 1 & 0 \\
0 & 1 & 0 & 0 \\
0 & 0 & 0 & 1
\end{bmatrix}.
\end{equation}
In this case, the $4 \times 4$ operation corresponds to just swapping 2 elements.
Hence, \cref{alg:2q-matvec1,alg:2q-matvec2,alg:2q-matvec3,alg:2q-matvec4} in \cref{alg:2q} yield
\begin{align*}
\psi[a_j] &= \phi[a_j], \\
\psi[b_j] &= \phi[c_j], \\
\psi[c_j] &= \phi[b_j], \\
\psi[d_j] &= \phi[d_j].
\end{align*}

We remark that \cref{alg:2q} allows to apply arbitrary 2-qubit gates to noncontiguous qubits as illustrated in \cref{fig:qgates}(e).

\subsection{Multi-qubit gates}
\label{sec:multi-q}

The algorithms for state vector simulation of 1- and 2-qubit gates can be generalized to multi-qubit gates in a straightforward manner.
For every additional qubit, we need to define 2 additional bit masks to efficiently calculate the indices via bit operations.

\begin{figure}[hbtp]
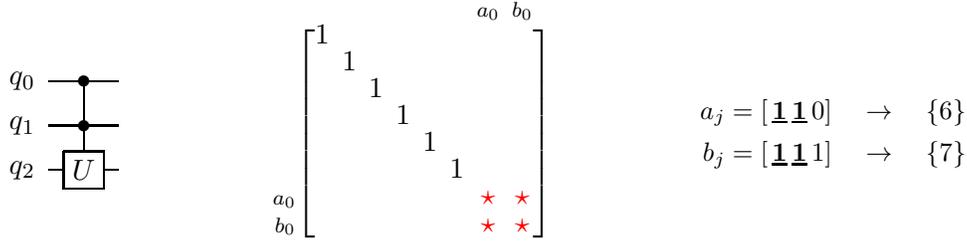

\centering
\begin{align*}
\quad{\begin{myqcircuit}
\lstick{q_0} & \ctrl1   & \qw \\
\lstick{q_1} & \ctrl1 & \qw \\
\lstick{q_2} & \gate{U}      & \qw
\end{myqcircuit}} &&
\setlength{\arraycolsep}{0.5ex}\renewcommand{\arraystretch}{0.75}
\begin{bNiceMatrix}[first-row,first-col]
      & & & & &  & & a_0 & b_0 \\
      & 1 & &  \\
      & & 1 & &  \\
      & & & 1 \\
      & & & & 1 \\
      & & & & & 1 \\
      & & & & & & 1 \\
  a_0 & & & & & & & \starC & \starC \\
  b_0 & & & & & & & \starC & \starC \\
\end{bNiceMatrix} &&
{\small\begin{aligned}
a_j &= [\, \underline{\textbf{1}} \, \underline{\textbf{1}} \, 0] \quad \to \quad \{ 6 \} \\
b_j &= [\, \underline{\textbf{1}} \, \underline{\textbf{1}} \, 1 ] \quad \to \quad \{ 7 \} \\
\end{aligned}}
\end{align*}
\caption{Graphical illustration of applying a double controlled 1-qubit gate $U$.}
\label{fig:dctrl-1q}
\end{figure}

However, the case of multi-controlled versions of smaller 1- or 2-qubit gates provides the most opportunity for performance gains. This is due to the fact that every additional control qubit halves the number of elements in the state vector that need to be updated.
For example, as illustrated in \cref{fig:dctrl-1q}, a doubly-controlled 1-qubit gate only updates 1/4 of the elements in the state vector. The Toffoli gate, or doubly-controlled NOT gate is the most common multi-qubit gate. Its implementation can be further optimized by swapping the elements of the state vector it acts on.

\section{Implementation details}
\label{sec:impl}

\qclabpp\ is an object-oriented, fully templated \cpp\ package for creating and representing quantum circuits.
\qclabpp\ can be used for rapid prototyping and testing of quantum algorithms, and allows for fast algorithm development and discovery.
\qclabpp\ has no external dependencies and provides I/O through openQASM making it compatible with quantum hardware.

In order to keep \qclabpp\ light-weighted and with a high degree of portability, all parallelization (both CPU and GPU) is done through \openmp.
Another advantage of \openmp\ GPU offloading is that the CPU and GPU implementations of the algorithms in \cref{sec:eff} are exactly the same except for the \openmp\ \texttt{\#pragma} statements.
This is illustrated in \cref{fig:Xgate-cpu-vs-gpu} which shows the CPU and GPU implementations for applying the Pauli-$X$ gate.
Note that both implementations only differ in the \texttt{\#pragma} statement above the for loop and the remaining part of the code is identical.

\begin{figure}[hbtp]
\centering
\begin{minipage}{0.5\textwidth}
\centering
\begin{lstlisting}[language=C++,basicstyle=\ttfamily\scriptsize,keywordstyle=\color{blue},commentstyle=\color{green!70!black},emph={pragma,omp,parallel},emphstyle=\color{red}]
template <typename T>
void apply(int nbQubits, int qubit, T* x) {

  // number of iterations
  int jmax = 1 << (nbQubits-1);

  // bit masks
  int mR = (1 << (nbQubits-qubit-1)) - 1;
  int mL = (1 << (nbQubits-1)) - 1 - mR;

  // loop over 2x2 matvecs
  #pragma omp parallel for
  for (int j = 0; j < jmax; j++) {
    int aj = (j & mR) + ((j & mL) << 1);
    int bj = aj + 1 << (nbQubits-qubit-1);
    std::swap(x[aj], x[bj]);
  }

}
\end{lstlisting}%
\end{minipage}%
\hfill%
\begin{minipage}{0.5\textwidth}
\centering
\begin{lstlisting}[language=C++,basicstyle=\ttfamily\scriptsize,keywordstyle=\color{blue},commentstyle=\color{green!70!black},emph={pragma,omp,parallel,target,teams,distribute},emphstyle=\color{red}]
template <typename T>
void apply_device(int nbQubits, int qubit, T* x) {

  // number of iterations
  int jmax = 1 << (nbQubits-1);

  // bit masks
  int mR = (1 << (nbQubits-qubit-1)) - 1;
  int mL = (1 << (nbQubits-1)) - 1 - mR;

  // loop over 2x2 matvecs
  #pragma omp target teams distribute parallel for
  for (int j = 0; j < jmax; j++) {
    int aj = (j & mR) + ((j & mL) << 1);
    int bj = aj + 1 << (nbQubits-qubit-1);
    std::swap(x[aj], x[bj]);
  }

}
\end{lstlisting}
\end{minipage}%
\caption{CPU (\emph{left}) versus GPU (\emph{right}) implementation  of the Pauli-$X$ gate in \qclabpp. Remark that they only differ in the \texttt{\#pragma} statement above the for loop.\label{fig:Xgate-cpu-vs-gpu}}
\end{figure}

\section{Numerical experiments}
\label{sec:exp}

All experiments presented in this section where performed on the NERSC Perlmutter supercomputer. Each GPU node in Perlmutter is equipped with 4 NVIDIA A100 40GB GPUs and an AMD EPYC 7763 CPU. The nodes are connected through a HPE Slingshot-11 interconnect. A single GPU has enough memory to store a 31 qubit state vector in complex double precision or a 32 qubit state vector in complex single precision. We first compare the \qclabpp\ CPU implementation to its GPU implementation. Next, we benchmark \qclabpp\ to other existing GPU-accelerated state vector simulation packages, such as cirq-qsim with the cuQuantum backend, and qibo both with the cupy and cuQuantum backends.

\begin{figure}[hbtp]
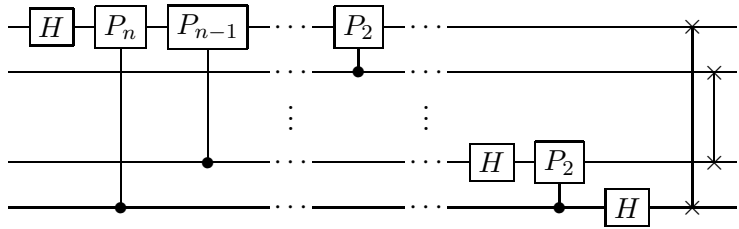
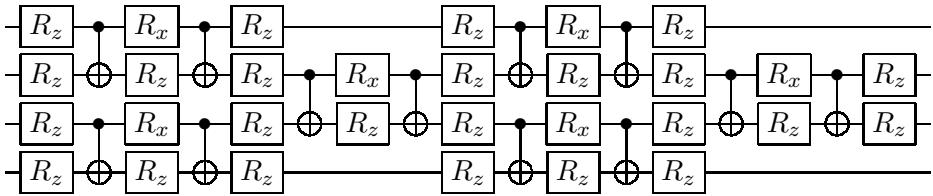

\centering
\begin{subfigure}{\textwidth}
\[
{\begin{myqcircuit*}{0.05}{0.75}
& \gate{H} & \gate{P_{n}} & \gate{P_{n-1}} & \qw & \cdots & & \gate{P_2} &  \qw & \cdots & & \qw & \qw & \qw & \qw & \qswap & \qw & \qw \\
& \qw & \qw & \qw  & \qw  & \cdots & & \ctrl{-1} & \qw & \cdots & & \qw & \qw & \qw & \qw & \qw & \qswap & \qw \\
& & & & & \raisebox{.5em}{\vdots} & & & &  \raisebox{.5em}{\vdots}  \\
& \qw & \qw & \ctrl{-3} & \qw  & \cdots & & \qw & \qw & \cdots & &\gate{H} & \gate{P_2} & \qw & \qw & \qw & \qswap \qwx[-2] & \qw \\
& \qw & \ctrl{-4} & \qw & \qw & \cdots & & \qw & \qw & \cdots & & \qw & \ctrl{-1} & \gate{H} & \qw & \qswap \qwx[-4] & \qw & \qw
\end{myqcircuit*}}
\]
\caption{QFT circuit}
\end{subfigure} \\[15pt]
\begin{subfigure}{\textwidth}
\[
{\begin{myqcircuit}
& \gate{R_z} & \ctrl1 & \gate{R_x} & \ctrl1 & \gate{R_z} & \qw & \qw & \qw & \gate{R_z} & \ctrl1 & \gate{R_x} & \ctrl1 & \gate{R_z} & \qw & \qw & \qw & \qw & \qw\\
& \gate{R_z} & \targ  & \gate{R_z} & \targ  & \gate{R_z} & \ctrl1 & \gate{R_x} & \ctrl1 & \gate{R_z} & \targ & \gate{R_z} & \targ & \gate{R_z} & \ctrl1 & \gate{R_x} & \ctrl1 & \gate{R_z} & \qw\\
& \gate{R_z} & \ctrl1 & \gate{R_x} & \ctrl1 & \gate{R_z} & \targ & \gate{R_z} & \targ & \gate{R_z} & \ctrl1 & \gate{R_x} & \ctrl1 & \gate{R_z} & \targ & \gate{R_z} & \targ & \gate{R_z} & \qw\\
& \gate{R_z} & \targ  & \gate{R_z} & \targ  & \gate{R_z} & \qw & \qw & \qw & \gate{R_z} & \targ & \gate{R_z} & \targ & \gate{R_z} & \qw & \qw & \qw & \qw & \qw
\end{myqcircuit}}
\]
\caption{Hamiltonian simulation circuit}
\end{subfigure}
\caption{Overview of the circuits used to benchmark the various quantum circuit simulators.}
\label{fig:bm-circ}
\end{figure}

We select two scalable classes of benchmarking circuits to run our experiments on, see \cref{fig:bm-circ}.
The first is the QFT circuit, which consists of single qubit Hadamard gates, controlled phase gates that connect every pair of qubits in the circuit, and ends with a series of SWAP gates.
The second circuit is a Trotter time evolution circuit for a 1D nearest-neighbor TFXY spin chain. This circuit consist of layers of single qubit rotations interleaved with leayers of CNOT gates acting on nearest-neighbor qubits. 

\subsection{\qclabpp: CPU versus GPU}
\label{sec:exp-cpu-vs-gpu}

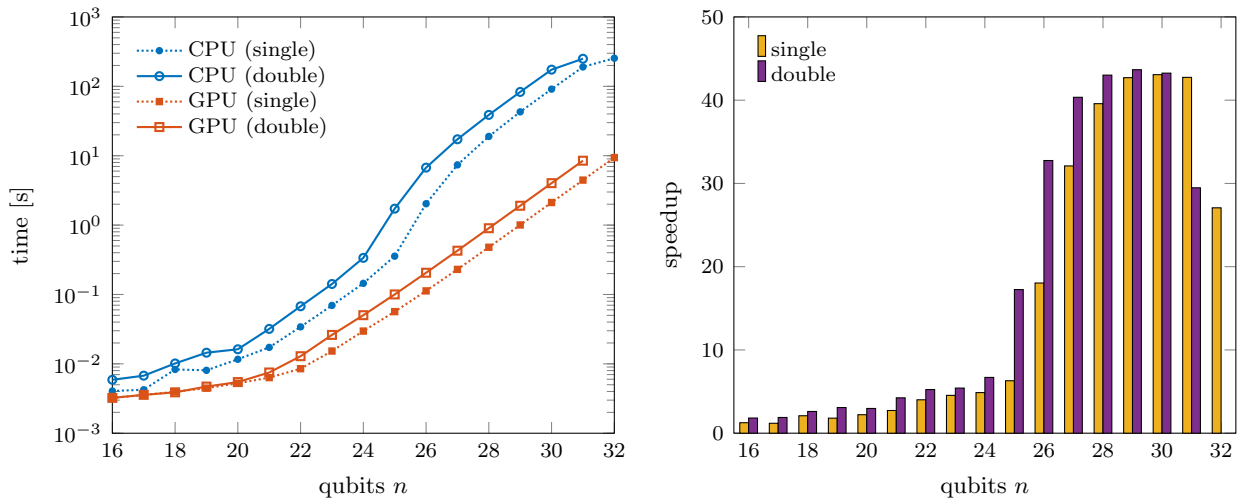
\begin{figure}[hbtp]
\centering
\begin{tikzpicture}
\begin{semilogyaxis}[
  width=0.5\textwidth,%
  xmin=16,xmax=32,%
  ymin=1e-3,ymax=1e3,%
  xtick={16,18,...,32},%
  ytick={1e-3,1e-2,1e-1,1e0,1e1,1e2,1e3},%
  xlabel={qubits $n$},%
  ylabel={time [s]},%
  legend style={draw=none,fill=none,row sep=-2pt},%
  legend pos=north west,%
]
\addplot[myColOne,thick,densely dotted,mark=*,mark size=1,mark options=solid]%
   table[x index=0,y index=1] {\datfile{qft-qclab-single}};
\addplot[myColOne,thick,mark=o,mark size=1.5]%
   table[x index=0,y index=1] {\datfile{qft-qclab-double}};
\addplot[myColTwo,thick,densely dotted,mark=square*,mark size=1,mark options=solid]%
   table[x index=0,y index=2] {\datfile{qft-qclab-single}};
\addplot[myColTwo,thick,mark=square,mark size=1.5]%
   table[x index=0,y index=2] {\datfile{qft-qclab-double}};
\legend{CPU (single),CPU (double),GPU (single),GPU (double)};
\end{semilogyaxis}
\end{tikzpicture}%
\hfill%
\begin{tikzpicture}
\begin{axis}[
  width=0.5\textwidth,%
  xmin=15.5,xmax=32.5,%
  ymin=0,ymax=50,%
  xtick={16,18,...,32},%
  ytick={0,10,...,50},%
  xlabel={qubits $n$},%
  ylabel={speedup},%
  bar width=0.3,%
  legend entries={single,double},%
  legend style={draw=none,fill=none,row sep=-2pt},%
  legend pos=north west,%
  legend image code/.code={\draw (0mm,-0.75mm) rectangle (1mm,2mm);},%
]
\addplot[ybar,fill=myorange,bar shift=-0.15]
   table[x index=0,y index=3] {\datfile{qft-qclab-single}};
\addplot[ybar,fill=mypurple,bar shift=0.15]
   table[x index=0,y index=3] {\datfile{qft-qclab-double}};
\end{axis}
\end{tikzpicture}
\caption{\qclabpp: CPU versus GPU for QFT circuit (Perlmutter).}
\label{fig:qclab-cpu-vs-gpu-qft}
\end{figure}

\begin{figure}[hbtp]
\centering
\begin{tikzpicture}
\begin{semilogyaxis}[
  width=0.5\textwidth,%
  xmin=16,xmax=32,%
  ymin=1e-2,ymax=1e4,%
  xtick={16,18,...,32},%
  ytick={1e-2,1e-1,1e0,1e1,1e2,1e3,1e4},%
  xlabel={qubits $n$},%
  ylabel={time [s]},%
  legend style={draw=none,fill=none,row sep=-2pt},%
  legend pos=north west,%
]
\addplot[myColOne,thick,densely dotted,mark=*,mark size=1,mark options=solid]%
   table[x index=0,y index=1] {\datfile{trotter-qclab-single}};
\addplot[myColOne,thick,mark=o,mark size=1.5]%
   table[x index=0,y index=1] {\datfile{trotter-qclab-double}};
\addplot[myColTwo,thick,densely dotted,mark=square*,mark size=1,mark options=solid]%
   table[x index=0,y index=2] {\datfile{trotter-qclab-single}};
\addplot[myColTwo,thick,mark=square,mark size=1.5]%
   table[x index=0,y index=2] {\datfile{trotter-qclab-double}};
\legend{CPU (single),CPU (double),GPU (single),GPU (double)};
\end{semilogyaxis}
\end{tikzpicture}%
\hfill%
\begin{tikzpicture}
\begin{axis}[
  width=0.5\textwidth,%
  xmin=15.5,xmax=32.5,%
  ymin=0,ymax=50,%
  xtick={16,18,...,32},%
  ytick={0,10,...,50},%
  xlabel={qubits $n$},%
  ylabel={speedup},%
  bar width=0.3,%
  legend entries={single,double},%
  legend style={draw=none,fill=none,row sep=-2pt},%
  legend pos=north west,%
  legend image code/.code={\draw (0mm,-0.75mm) rectangle (1mm,2mm);},%
]
\addplot[ybar,fill=myorange,bar shift=-0.15]
   table[x index=0,y index=3] {\datfile{trotter-qclab-single}};
\addplot[ybar,fill=mypurple,bar shift=0.15]
   table[x index=0,y index=3] {\datfile{trotter-qclab-double}};
\end{axis}
\end{tikzpicture}
\caption{\qclabpp: CPU versus GPU for Hamiltonian simulation circuit (Perlmutter).}
\label{fig:qclab-cpu-vs-gpu-ham}
\end{figure}
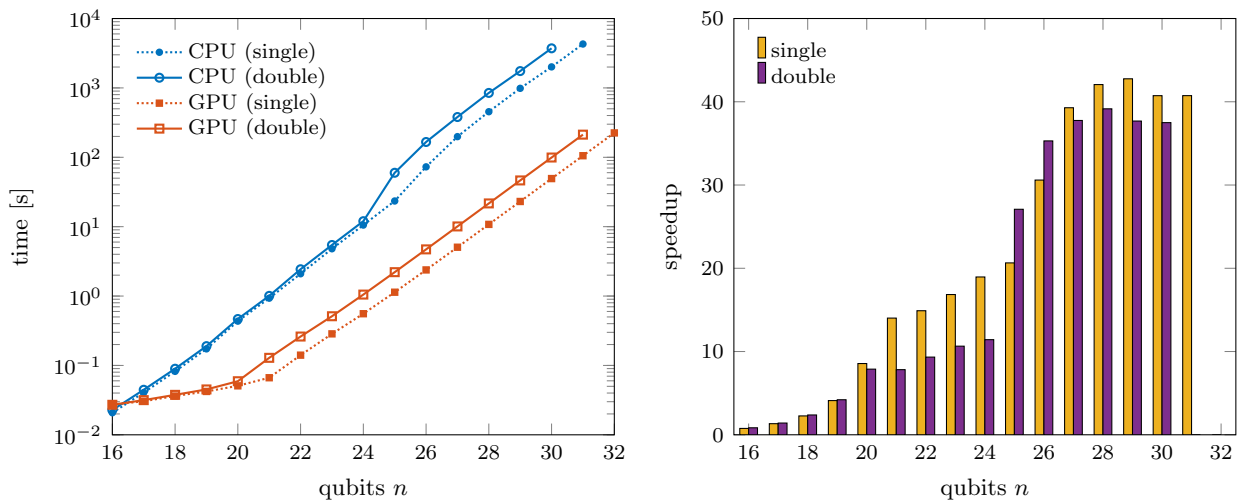

In a first experiment, we compare the CPU and GPU implementations in \qclabpp.
The timinig results for the QFT and Hamiltonian simulation circuits are presented in \cref{fig:qclab-cpu-vs-gpu-qft,fig:qclab-cpu-vs-gpu-ham}, respectively.
Both figures show timings and speedup factors for single and double precision.

We observe that the GPU kernels exhibit a perfect linear scaling on the loglog plot for systems with more than 22 qubits. Every additional qubit doubles the simulation time. The CPU simulation exhibits less regular scaling in the timings, likely due to memory access effects such as NUMA domains.
Furthermore, the GPU simulation significantly outperforms the CPU implementation leading to speedup factors of over $40\times$.
Note that the results are qualitatively comparable for both sets of benchmarking circuits.
The main difference is that the Hamiltonian simulation circuit shows larger GPU speedups in the 19-25 qubit range compared to the QFT circuit.

\subsection{GPU benchmarking}
\label{sec:exp-gpu}

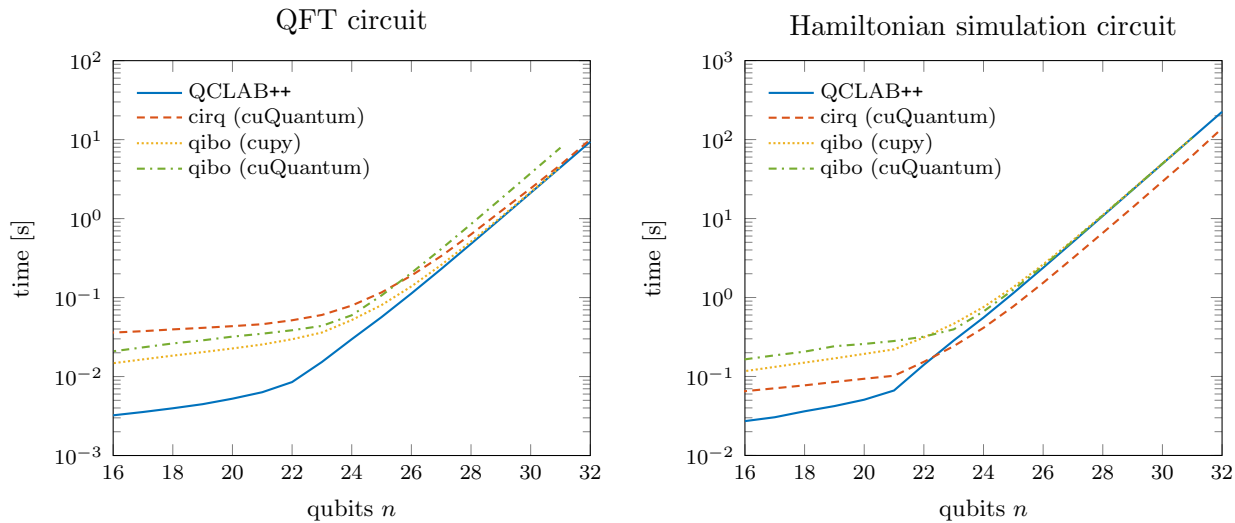
\begin{figure}[hbtp]
\centering
\begin{tikzpicture}
\begin{semilogyaxis}[
  width=0.48\textwidth,%
  xmin=16,xmax=32,%
  ymin=1e-3,ymax=1e2,%
  xtick={16,18,...,32},%
  ytick={1e-3,1e-2,1e-1,1e0,1e1,1e2},%
  xlabel={qubits $n$},%
  ylabel={time [s]},%
  title={QFT circuit},%
  legend style={draw=none,fill=none,row sep=-2pt},%
  legend pos=north west,%
]
\addplot[myColOne,thick]%
   table[x index=0,y index=2] {\datfile{qft-qclab-single}};
\addplot[myColTwo,thick,densely dashed]%
   table[x index=0,y index=1] {\datfile{qft-cirq-cuquantum-single}};
\addplot[myColThr,thick,densely dotted]%
   table[x index=0,y index=1] {\datfile{qft-qibo-cupy-single}};
\addplot[myColFiv,thick,dash dot]%
   table[x index=0,y index=1] {\datfile{qft-qibo-cuquantum-single}};
\legend{QCLAB\texttt{++},cirq (cuQuantum),qibo (cupy),qibo (cuQuantum)};
\end{semilogyaxis}
\end{tikzpicture}%
\hfill%
\begin{tikzpicture}
\begin{semilogyaxis}[
  width=0.48\textwidth,%
  xmin=16,xmax=32,%
  ymin=1e-2,ymax=1e3,%
  xtick={16,18,...,32},%
  ytick={1e-2,1e-1,1e0,1e1,1e2,1e3},%
  xlabel={qubits $n$},%
  ylabel={time [s]},%
  title={Hamiltonian simulation circuit},%
  legend style={draw=none,fill=none,row sep=-2pt},%
  legend pos=north west,%
]
\addplot[myColOne,thick]%
   table[x index=0,y index=2] {\datfile{trotter-qclab-single}};
\addplot[myColTwo,thick,densely dashed]%
   table[x index=0,y index=1] {\datfile{trotter-cirq-cuquantum-single}};
\addplot[myColThr,thick,densely dotted]%
   table[x index=0,y index=1] {\datfile{trotter-qibo-cupy-single}};
\addplot[myColFiv,thick,dash dot]%
   table[x index=0,y index=1] {\datfile{trotter-qibo-cuquantum-single}};
\legend{QCLAB\texttt{++},cirq (cuQuantum),qibo (cupy),qibo (cuQuantum)};
\end{semilogyaxis}
\end{tikzpicture}
\caption{GPU benchmarking on Perlmutter (single precision).}
\label{fig:gpu-qft-ham-single}
\end{figure}

In our second experiment, we compare the GPU state vector simulator of \qclabpp\ to the cuQuantum backend provided by cirq-qsim and to qibo, where we use both the cupy and cuQuantum backends. The timings for the single precision simulations are summarized in \cref{fig:gpu-qft-ham-single}.
The most notable difference appears in the smaller qubit regime, 16-25 qubits for the QFT circuit and 16-21 qubits for the Hamiltonian simulation circuit, where \qclabpp\ shows a signficant advantage over the other simulators.
The most likely reason for this difference is that \qclabpp\ runs a compiled \cpp\ code which has a lower overhead compared to the Python interfaces the other simulators use.
At the larger qubit sizes, the cost of the Python overhead becomes negligible and all simulators deliver a similar performance. We conclude that \qclabpp\ is competitive with other simulators while offering some distinct advantages.

\section*{Acknowledgements}
This research used resources of the National Energy Research Scientific Computing Center, a DOE Office of Science User Facility supported by the Office of Science of the U.S. Department of Energy under Contract No.~DE-AC02-05CH11231 using NERSC award ASCR-ERCAP0024463.

\bibliographystyle{abbrvurl}
\bibliography{references}

\begin{thebibliography}{10}

\bibitem{pennylane}
V.~Bergholm et~al.
\newblock Pennylane: Automatic differentiation of hybrid quantum-classical
  computations, 2018.
\newblock \href {https://doi.org/10.48550/arXiv.1811.04968}
  {\path{doi:10.48550/arXiv.1811.04968}}.

\bibitem{qclab}
D.~Camps and R.~{Van Beeumen}.
\newblock Quantumcomputinglab/qclab: Qclab v0.1.2, Aug. 2021.
\newblock \href {https://doi.org/10.5281/zenodo.5160555}
  {\path{doi:10.5281/zenodo.5160555}}.

\bibitem{qft_camps}
D.~Camps, R.~Van~Beeumen, and C.~Yang.
\newblock Quantum fourier transform revisited.
\newblock {\em Numerical Linear Algebra with Applications}, 28(1):e2331, 2021.
\newblock \href {https://doi.org/10.1002/nla.2331}
  {\path{doi:10.1002/nla.2331}}.

\bibitem{cirq}
{Cirq Developers}.
\newblock Cirq, Dec. 2022.
\newblock {See full list of authors on Github: https://github
  .com/quantumlib/Cirq/graphs/contributors}.
\newblock \href {https://doi.org/10.5281/zenodo.7465577}
  {\path{doi:10.5281/zenodo.7465577}}.

\bibitem{qft}
D.~Coppersmith.
\newblock An approximate fourier transform useful in quantum factoring, 1994.

\bibitem{qibojit_paper}
S.~Efthymiou, M.~Lazzarin, A.~Pasquale, and S.~Carrazza.
\newblock Quantum simulation with just-in-time compilation.
\newblock {\em {Quantum}}, 6:814, Sept. 2022.
\newblock \href {https://doi.org/10.22331/q-2022-09-22-814}
  {\path{doi:10.22331/q-2022-09-22-814}}.

\bibitem{qibo_paper}
S.~Efthymiou, S.~Ramos-Calderer, C.~Bravo-Prieto, A.~P{\'{e}}rez-Salinas,
  D.~Garc{\'{\i}}a-Mart{\'{\i}}n, A.~Garcia-Saez, J.~I. Latorre, and
  S.~Carrazza.
\newblock Qibo: a framework for quantum simulation with hardware acceleration.
\newblock {\em Quantum Science and Technology}, 7(1):015018, dec 2021.
\newblock \href {https://doi.org/10.1088/2058-9565/ac39f5}
  {\path{doi:10.1088/2058-9565/ac39f5}}.

\bibitem{cuquantum}
L.~Fang, ahehn nv, hhbayraktar, and sam stanwyck.
\newblock Nvidia/cuquantum: cuquantum python v22.11.0.1, Jan. 2023.
\newblock \href {https://doi.org/10.5281/zenodo.7523366}
  {\path{doi:10.5281/zenodo.7523366}}.

\bibitem{quest}
T.~Jones, A.~Brown, I.~Bush, and S.~C. Benjamin.
\newblock {QuEST} and high performance simulation of quantum computers.
\newblock {\em Scientific Reports}, 9(1), July 2019.
\newblock \href {https://doi.org/10.1038/s41598-019-47174-9}
  {\path{doi:10.1038/s41598-019-47174-9}}.

\bibitem{Liu2022}
H.~Liu, G.~H. Low, D.~S. Steiger, T.~H\"{a}ner, M.~Reiher, and M.~Troyer.
\newblock Prospects of quantum computing for molecular sciences.
\newblock {\em Materials Theory}, 6(1), Mar. 2022.
\newblock \href {https://doi.org/10.1186/s41313-021-00039-z}
  {\path{doi:10.1186/s41313-021-00039-z}}.

\bibitem{Qiskit}
{Qiskit Developers}.
\newblock Qiskit: An open-source framework for quantum computing, 2021.
\newblock \href {https://doi.org/10.5281/zenodo.2573505}
  {\path{doi:10.5281/zenodo.2573505}}.

\bibitem{GoogleErrorCorrection}
{Quantum AI team}.
\newblock Suppressing quantum errors by scaling a surface code logical qubit.
\newblock {\em Nature}, 614(7949):676--681, Feb. 2023.
\newblock \href {https://doi.org/10.1038/s41586-022-05434-1}
  {\path{doi:10.1038/s41586-022-05434-1}}.

\bibitem{qsim}
{Quantum AI team and collaborators}.
\newblock qsim, Sept. 2020.
\newblock \href {https://doi.org/10.5281/zenodo.4023103}
  {\path{doi:10.5281/zenodo.4023103}}.

\bibitem{QuantinuumErrorSuppression}
C.~Ryan-Anderson et~al.
\newblock Implementing fault-tolerant entangling gates on the five-qubit code
  and the color code, 2022.
\newblock \href {https://doi.org/10.48550/arXiv.2208.01863}
  {\path{doi:10.48550/arXiv.2208.01863}}.

\bibitem{qclabpp}
R.~{Van Beeumen} and D.~Camps.
\newblock Quantumcomputinglab/qclabpp: Qclab++ v0.1.2, Aug. 2021.
\newblock \href {https://doi.org/10.5281/zenodo.5160682}
  {\path{doi:10.5281/zenodo.5160682}}.

\bibitem{Williams2022}
H.~J. Williams.
\newblock Versatile neutral atoms take on quantum circuits.
\newblock {\em Nature}, 604(7906):429--430, Apr. 2022.
\newblock \href {https://doi.org/10.1038/d41586-022-01029-y}
  {\path{doi:10.1038/d41586-022-01029-y}}.

\end{thebibliography}

\end{document}